\documentclass{PoS}
\usepackage{amsmath}

\newcommand{\al}{\ensuremath{\alpha} }
\newcommand{\be}{\ensuremath{\beta} }
\newcommand{\ga}{\ensuremath{\gamma} }
\newcommand{\eps}{\ensuremath{\epsilon} }
\newcommand{\la}{\ensuremath{\lambda} }
\newcommand{\si}{\ensuremath{\sigma} }
\newcommand{\Si}{\ensuremath{\Sigma} }
\newcommand{\om}{\ensuremath{\omega} }
\newcommand{\psibar}{\ensuremath{\overline\psi} }
\newcommand{\gsim}{\ensuremath{\gtrsim} }
\newcommand{\X}{\ensuremath{\!\times\!} }
\newcommand{\ReTr}[1]{\ensuremath{\mbox{ReTr}\left[ #1 \right]} }
\newcommand{\vev}[1]{\ensuremath{\left\langle #1 \right\rangle} }
\newcommand{\pbp}{\ensuremath{\vev{\psibar\psi}} }
\newcommand{\eq}[1]{Eq.~\ref{#1}}
\newcommand{\fig}[1]{Fig.~\ref{#1}}
\newcommand{\refcite}[1]{Ref.~\cite{#1}}
\newcommand{\secref}[1]{Section~\ref{#1}}
\newcommand{\mysection}[1]{\vspace{-6 pt}\section{#1}\vspace{-6 pt}}
\newcommand{\mutesection}[1]{\vspace{-6 pt}\section*{#1}\vspace{-6 pt}}

\title{Eight light flavors on large lattice volumes}
\ShortTitle{Eight light flavors on large lattice volumes}

\author{\speaker{David Schaich} for USBSM \\
  Department of Physics, University of Colorado, Boulder, CO 80309, United States \\
  Department of Physics, Syracuse University, Syracuse, NY 13244, United States\thanks{Present address} \\
  E-mail: \email{dschaich@syr.edu}
}

\abstract{
  I present first results from large-scale lattice investigations of SU(3) gauge theory with eight light flavors in the fundamental representation.
  Using leadership computing resources at Argonne, we are generating gauge configurations with lattice volumes up to $64^3\X128$ at relatively strong coupling, in an attempt to access the chiral regime. 
  We use nHYP-improved staggered fermions, carefully monitoring finite-volume effects and other systematics.
  Here I focus on analyses of the light hadron spectrum and chiral condensate, measured on lattice volumes up to $48^3\X96$ with fermion masses as light as $m = 0.004$ in lattice units.
  We find no clear indication of spontaneous chiral symmetry breaking in these observables.
  I discuss the implications of these initial results, and prospects for further physics projects employing these ensembles of gauge configurations.
}

\FullConference{31st International Symposium on Lattice Field Theory \\ 29 July -- 3 August 2013 \\ Mainz, Germany}

\begin{document}
\setlength{\abovedisplayskip}{6 pt}
\setlength{\belowdisplayskip}{6 pt}
SU(3) gauge theory with $N_f = 8$ fermions in the fundamental representation is a very interesting system to investigate through non-perturbative lattice calculations.
Initial lattice studies of this theory explored its chiral dynamics, observing no indications of IR conformality and concluding that the system most likely undergoes spontaneous chiral symmetry breaking~\cite{Appelquist:2007hu, Deuzeman:2008sc, Fodor:2009wk, Hasenfratz:2010fi, Jin:2010vm}.
More recent investigations have added to this picture, with \refcite{Cheng:2013eu} reporting a large mass anomalous dimension across a wide range of energy scales, and \refcite{Aoki:2013xza} arguing that the eight-flavor theory exhibits some remnant of IR conformality despite chiral symmetry breaking. 
Finally, \refcite{Aoki:2013qxa} presents preliminary observations of a light flavor-singlet scalar in this system, with a scalar mass consistent with the 125~GeV Higgs boson (within large uncertainties). 

These results motivate further investment of computational resources to study eight light flavors on large lattice volumes.
Anticipating this need, the USQCD Collaboration\footnote{\href{http://www.usqcd.org}{http://www.usqcd.org}} made eight-flavor configuration generation one of its community projects running on the ``Intrepid'' Blue Gene/P system at the Argonne Leadership Computing Facility, through the INCITE program\footnote{\href{http://www.doeleadershipcomputing.org}{http://www.doeleadershipcomputing.org}} of the U.S.\ Department of Energy (DOE).
This is the first INCITE project managed by the ``USBSM'' community within USQCD,\footnote{\href{http://bsm.physics.yale.edu}{http://bsm.physics.yale.edu}} and this proceedings presents a first look at our progress and preliminary results.
Additional projects being pursued by USBSM are summarized in a recent white paper~\cite{Appelquist:2013sia}.

Assuming that the eight-flavor theory is chirally broken, the ultimate aim of this effort is to carry out controlled fits to chiral perturbation theory ($\chi$PT), and thereby determine low-energy constants (LECs) of the chiral lagrangian.
These LECs include the ratio $B / F \propto \pbp / F^3$ that governs the enhancement of the chiral condensate relative to the symmetry breaking scale $F$, the electroweak $S$ parameter, and even WW scattering parameters~\cite{Appelquist:2010xv, Appelquist:2012sm, Appelquist:2013sia}.
While this ambitious goal may not be achieved immediately, we are already performing many analyses using the ensembles being generated, which feature larger volumes and lighter masses than obtained previously.
Even our initial results provide valuable new information about the eight-flavor system, and allow us to assess the prospects for reaching the chiral regime.

I begin in the next section by reviewing our nHYP-improved staggered lattice action and the current status of configuration generation, including thermalization and auto-correlations.
\secref{sec:spectrum} presents some initial results for meson masses and the pseudoscalar decay constant, checking finite-volume effects and comparing linear chiral extrapolations with power-law fits.
In \secref{sec:pbp} I focus on the chiral condensate, comparing three different ways to investigate this quantity: from direct measurements, through leading-order $\chi$PT (the Gell-Mann--Oakes--Renner relation), and using the eigenvalues of the massless Dirac operator.
\secref{sec:conclusion} concludes by reviewing additional projects that will use these eight-flavor gauge configurations.
Because many of our ensembles are still being generated, the results below are all preliminary and subject to change before final publication.

\mysection{Lattice action and status of configuration generation} 
Many-flavor lattice calculations are carried out at stronger bare couplings than QCD, which requires appropriate improvement of the lattice action.
Our gauge action includes an adjoint plaquette term with coefficient $\be_A = -\be_F / 4$ (so that perturbatively $\be_F \simeq 12 / g^2$), and we use staggered fermions improved with one step of nHYP smearing.
The nHYP smearing parameters are $\al = (0.5, 0.5, 0.4)$, small enough to avoid numerical instabilities.
Taking advantage of previous studies of this action~\cite{Cheng:2011ic, Hasenfratz:2013uha}, we focus our attention on $\be_F = 5.0$, a relatively strong coupling that is still weak enough to avoid a lattice phase.

We have implemented this action in QHMC (a.k.a.~FUEL, ``Framework for Unified Evolution of Lattices''), a USQCD software package currently under development.
QHMC is being designed to allow flexible experimentation with integration algorithms.
To generate large-volume configurations on Intrepid, we use an HMC algorithm with two to four Hasenbusch masses, depending on the volume and fermion mass. 
Subsequent measurements (e.g., of the Wilson flow, spectrum and eigenvalues) are carried out on clusters using code based in part on the MILC Collaboration's public lattice gauge theory software.\footnote{\href{http://www.physics.utah.edu/~detar/milc/}{http://www.physics.utah.edu/$\sim$detar/milc/}}

\begin{table}[t]
  \caption{\label{ensembles} Accumulated molecular dynamics time units (MDTU) for $N_f = 8$ ensembles.  Configuration generation is still underway for ensembles marked with $^{\dag}$.}
  \centering
  \begin{tabular}{|c||c|c|c|c|c|}
    \hline
    Fermion mass  & $64^3\X128$   & $48^3\X96$      & $32^3\X64$      & $24^3\X48$  & $16^3\X32$  \\\hline\hline
    0.002         &  70$^{\dag}$  &                 &                 &             &             \\\hline
    0.003         & 185$^{\dag}$  &                 &                 &             &             \\\hline
    0.004         & 252$^{\dag}$  & 1 032$^{\dag}$  &                 &             &             \\\hline
    0.006         &               & 1 110$^{\dag}$  &                 &             &             \\\hline
    0.008         &               & 1 630$^{\dag}$  & 3 024           &             &             \\\hline
    0.010         &               & 2 142$^{\dag}$  & 6 372$^{\dag}$  &  3 012      &             \\\hline
    0.015         &               &                 & 3 018           & 10 074      &             \\\hline
    0.020         &               &                 &                 & 10 074      &  3 000      \\\hline
    0.030         &               &                 &                 &             &  3 000      \\\hline
    0.040         &               &                 &                 &             &  3 000      \\\hline
    0.050         &               &                 &                 &             &  3 000      \\\hline
  \end{tabular}
\end{table}

Our volumes vary from $16^3\X32$ to $64^3\X128$, with overlapping ranges of fermion masses $m$ that we use to monitor finite-volume effects.
Table~\ref{ensembles} lists our ensembles, reporting the number of molecular dynamics time units (MDTU) accumulated as of October 2013.
The initial target for each ensemble is $\sim$3000~MDTU, to provide ample statistics after thermalization.
In addition, two $24^3\X48$ ensembles (with $m = 0.02$ and 0.015) and one $32^3\X64$ ensemble (with $m = 0.01$) are being extended to $\sim$10,000 MDTU for investigations of glueballs and the scalar spectrum.

\begin{figure}[b]
  \includegraphics[width=0.45\linewidth]{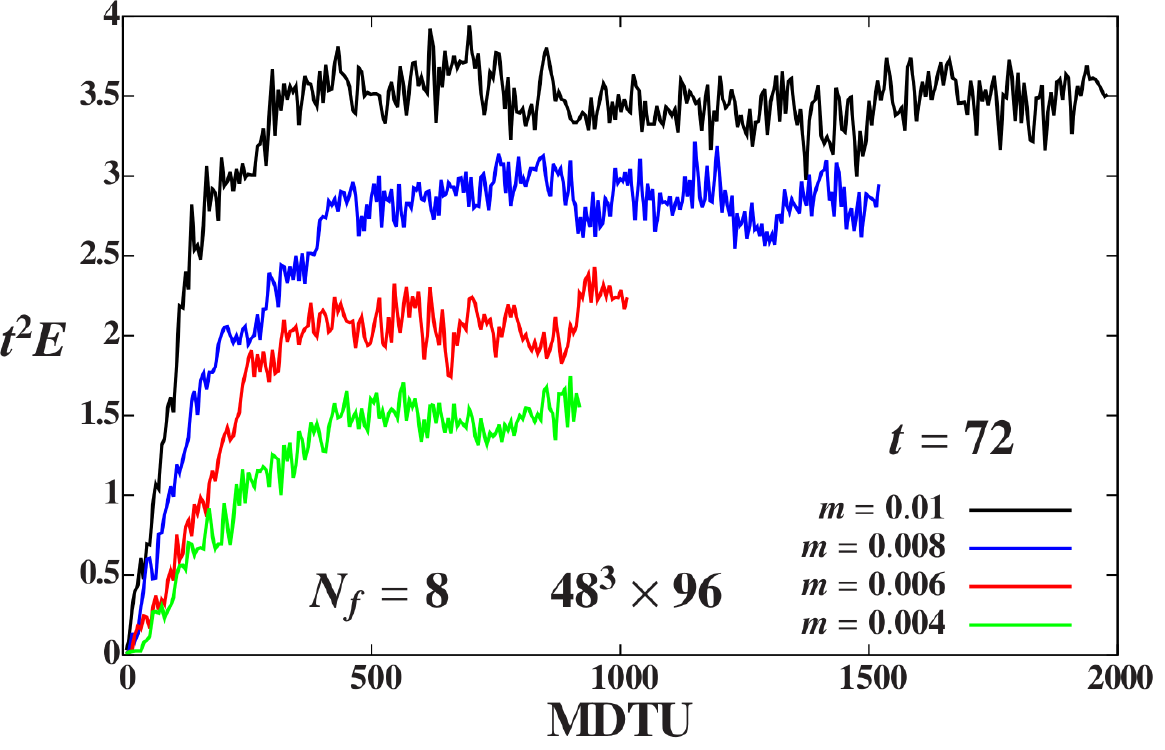}\hfill
  \includegraphics[width=0.45\linewidth]{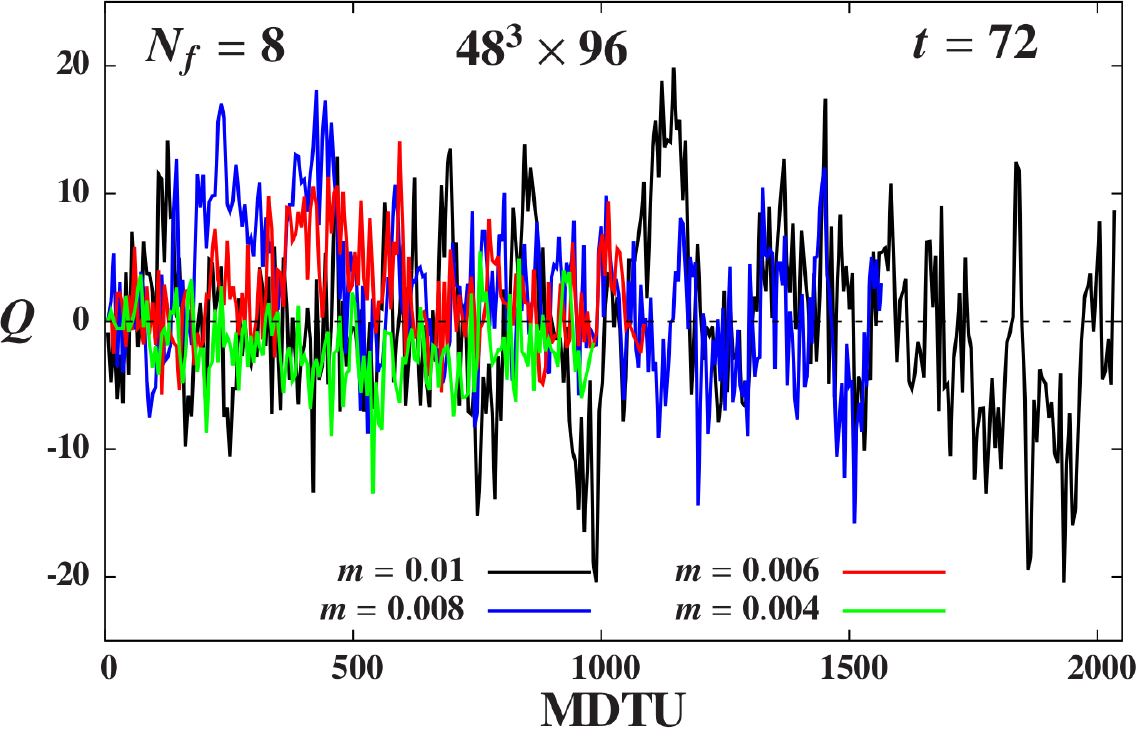}
  \caption{\label{fig:therm} Time-series plots of $t^2 E$ (left) and the topological charge (right) for $48^3\X96$ ensembles, from \protect\eq{eq:Wflow} after running the Wilson flow to $\sqrt{8t} = L / 2$.}
\end{figure}

Regarding thermalization, we have found Wilson flow observables measured after long flow times $t$ to be sensitive probes of thermalization (more so than $\pbp$, for example).
The Wilson flow integrates out to distances $\sim \sqrt{8t}$, providing relatively inexpensive long-range quantities.
\fig{fig:therm} shows representative time-series plots for the energy $E$ and topological charge $Q$,
\begin{align}
  \label{eq:Wflow}
  E & = -\frac{1}{2}\ReTr{F_{\mu\nu}F^{\mu\nu}} &
  Q & = \frac{1}{32\pi^2}\ReTr{\eps_{\mu\nu\rho\si} F^{\mu\nu}F^{\rho\si}},
\end{align}
both constructed from the clover-leaf definition of $F_{\mu\nu}$. 
Such time-series plots of $t^2 E$ indicate that most ensembles thermalize within 300 MDTU.
The lighter-mass $48^3\X96$ ensembles take longer to thermalize, 400 MDTU for $m = 0.008$ and 0.006, and 450 MDTU for $m = 0.004$.
We anticipate that the $64^3\X128$ ensembles will require $\sim$500 MDTU for thermalization.

The left panel of \fig{fig:therm} shows that the gauge fields evolve significantly from the initial configuration.
Subsequent auto-correlation times should therefore be shorter than the time for the runs to thermalize.
The right panel of \fig{fig:therm} reinforces this expectation, demonstrating healthy fluctuations in the topological charge from \eq{eq:Wflow}, with topological susceptibility $\chi_t$ roughly quadratic in $m$. 

\mysection{\label{sec:spectrum}Meson masses and the pseudoscalar decay constant} 
\begin{figure}[b]
  \includegraphics[width=0.45\linewidth]{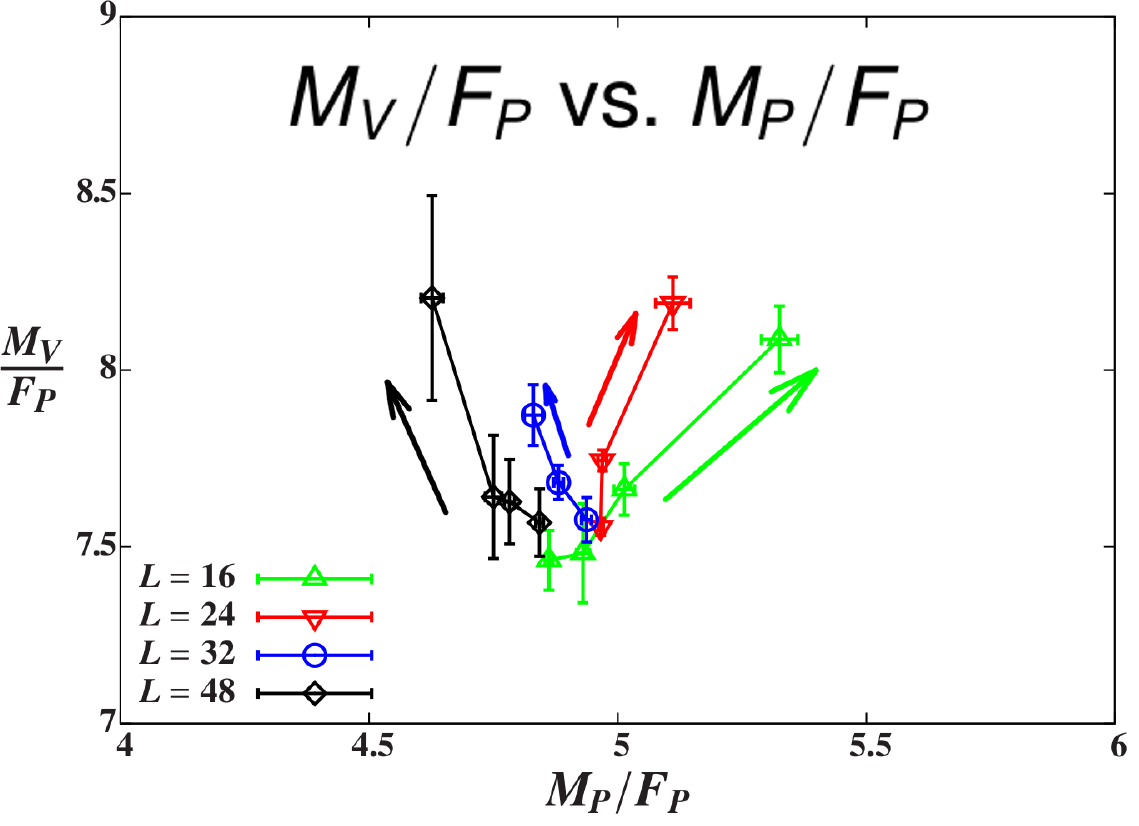}\hfill
  \includegraphics[width=0.45\linewidth]{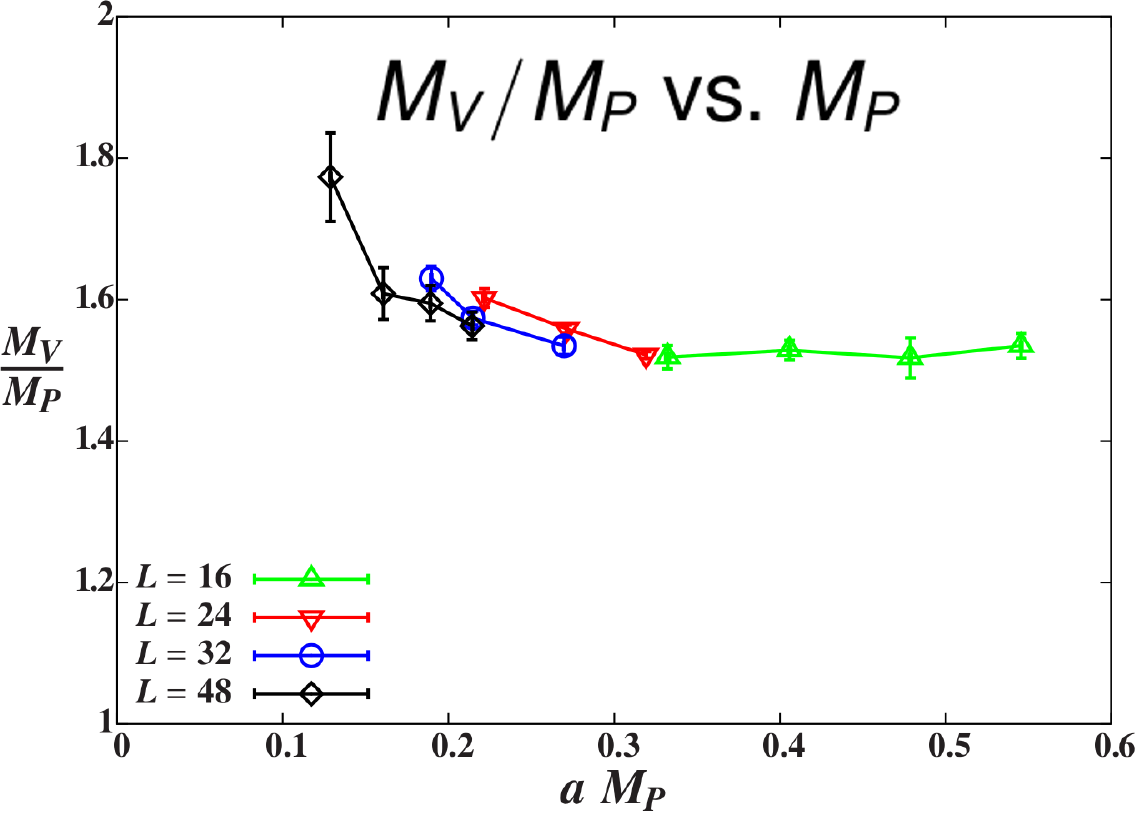}
  \caption{\label{fig:Edin} Left: Edinburgh-style plot of $M_V / F_P$ vs.\ $M_P / F_P$, with arrows indicating the direction of decreasing $m$.  Right: The ratio $M_V / M_P$ vs.\ $M_P$, which in a chirally broken system would diverge as $M_P \to 0$.}
\end{figure}

Let us first consider which of our ensembles may be subject to significant finite-volume effects.
The left panel of \fig{fig:Edin} plots $M_V / F_P$ vs.\ $M_P / F_P$, where $M_V$ ($M_P$) is the vector (pseudoscalar) meson mass and $F_P$ is the pseudoscalar decay constant.
These ratios are designed to emphasize finite-volume effects, which we expect to increase $M$ while decreasing $F_P$, pushing the points up and to the right.
Such behavior is clearly visible for the lightest $16^3\X32$ and $24^4\X48$ points, while the larger volumes move up and to the left for smaller $m$.
In systems exhibiting spontaneous chiral symmetry breaking (S$\chi$SB), we would expect these ratios to move the left as $m$ decreases, since $M_P \to 0$ while the other quantities remain non-zero.
We may be seeing initial signs of such behavior in \fig{fig:Edin}, though this is not yet clear, even for $m = 0.004$ on a $48^3\X96$ volume.

Another way of searching for S$\chi$SB in these data is to plot $M_V / M_P$ vs.\ $M_P$, as in the right panel of \fig{fig:Edin}.
If $M_V$ remains non-zero in the chiral limit, this ratio diverges as $M_P \to 0$.
Only the lightest $48^3\X96$ point shows a significant increase, ending up about 16\% larger than the heaviest $24^3\X48$ point.
This $48^3\X96$ ensemble with $m = 0.004$ is currently our shortest run, has our longest thermalization time, and may suffer from finite-volume effects.
The $64^3\X128$ runs in Table~\ref{ensembles} seem to be required in order to access the chiral regime.

\begin{figure}[b]
  \includegraphics[width=0.45\linewidth]{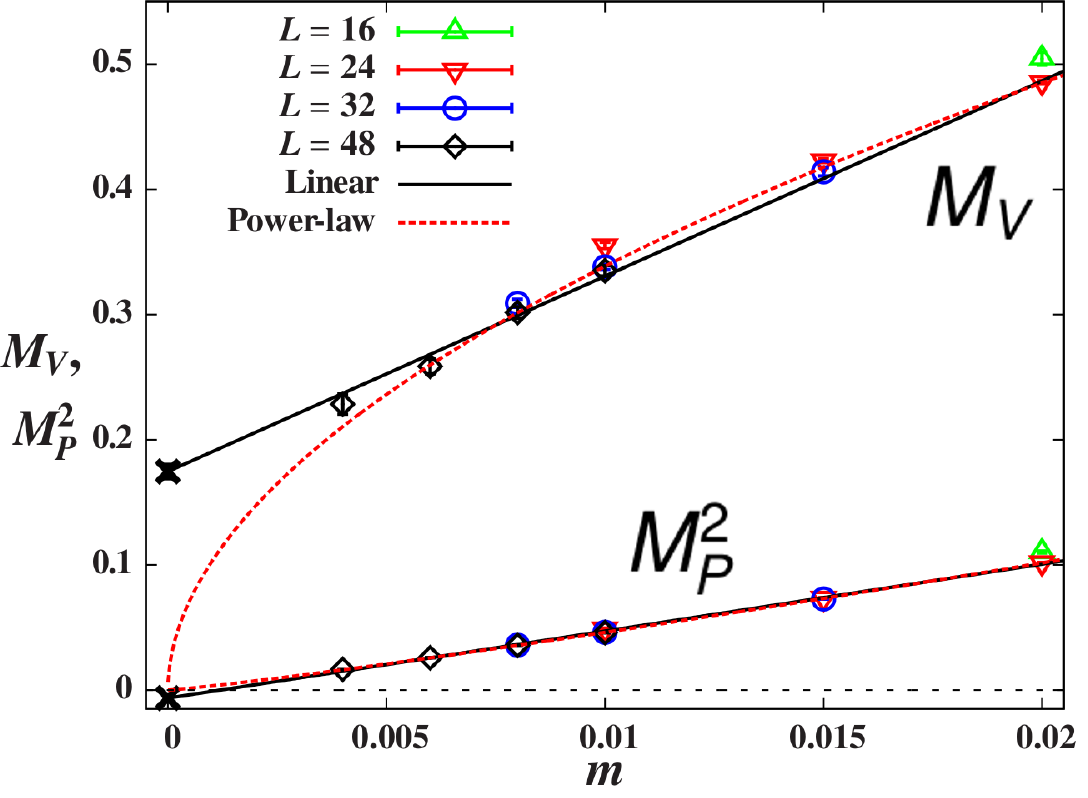}\hfill
  \includegraphics[width=0.45\linewidth]{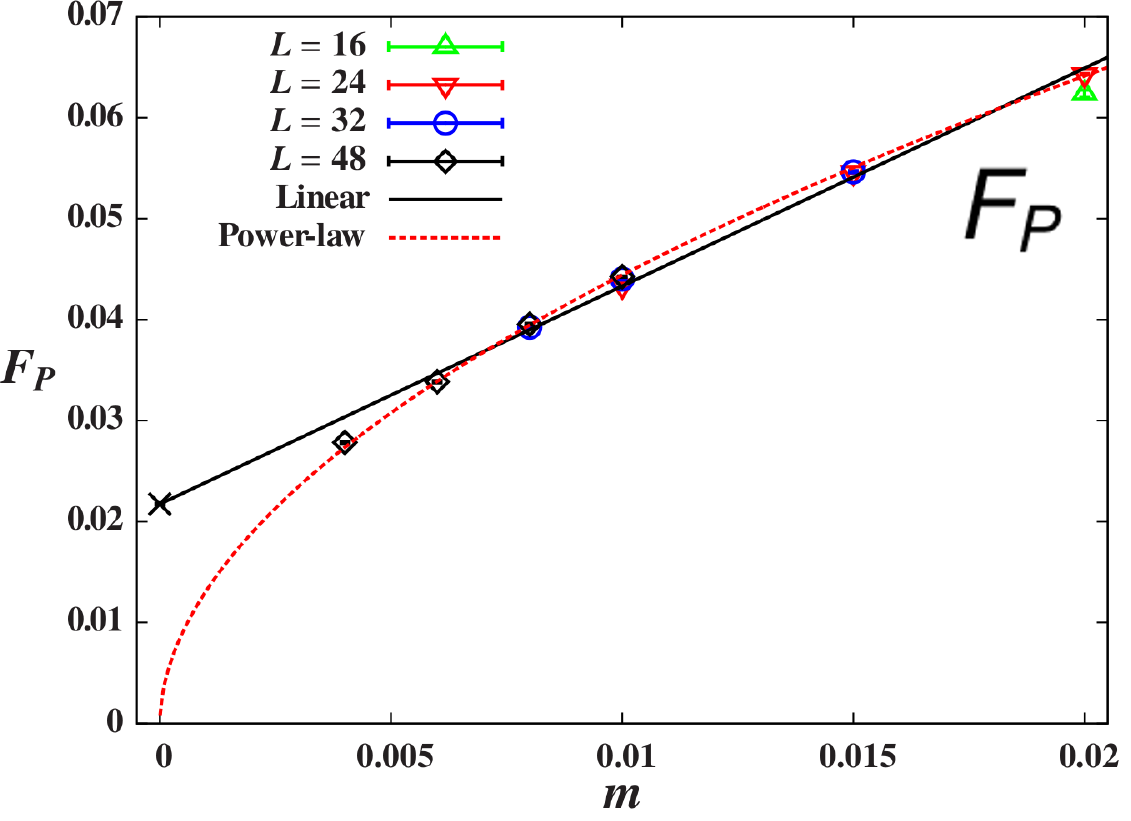}
  \caption{\label{fig:extrap} Linear and power-law chiral extrapolations for $M_P^2$ and $M_V$ (left) and $F_P$ (right).  All fits use the six points specified in \protect\eq{eq:fits}.}
\end{figure}

We can also see this by considering the right panel of \fig{fig:extrap}, which presents chiral extrapolations of $F_P$.
Simple linear extrapolations give $F \equiv \lim_{m \to 0} F_P = 0.021(2)$, so that $L \gsim 50$ is required to satisfy $FL \gsim 1$.
All linear and power-law fits shown in \fig{fig:extrap} use the following six points:
\begin{align}
  \label{eq:fits}
  0.004 \leq m \leq 0.01 & \mbox{ on } 48^3\X96 &
  m = 0.015 & \mbox{ on } 32^3\X64 &
  m = 0.02 & \mbox{ on } 24^3\X48.
\end{align}
By omitting either or both the lightest ($m = 0.004$) and heaviest ($m = 0.02$) data points, we observe significant dependence on the fit range, which dominates the current uncertainty on $F$.

The power-law fits have smaller $\chi^2$ than the linear extrapolations, typically by an order of magnitude or more, for the same number of degrees of freedom.
Writing the power as $M \propto m^{1 / (1 + \ga_m)}$, we find $\ga_m = 0.76(2)$ for $M_P$, while $\ga_m = 0.95(5)$ for $M_V$ and $\ga_m = 0.92(3)$ for $F_P$, again with uncertainties dominated by the fit range in $m$.
(As mentioned in the previous section, $\chi_t \propto m^{4 / (1 + \ga_m)}$ with $\ga_m = 0.76(10)$.)
Finite-size scaling, using all 14 thermalized ensembles in Table~\ref{ensembles}, prefers slightly smaller values: $\ga_m = 0.732(2)$ for $M_P$, $\ga_m = 0.86(2)$ for $M_V$ and $\ga_m = 0.779(3)$ for $F_P$.

\mysection{\label{sec:pbp}Chiral condensate: Direct measurements, GMOR relation and Dirac eigenvalues} 
\begin{figure}[b]
  \includegraphics[width=0.45\linewidth]{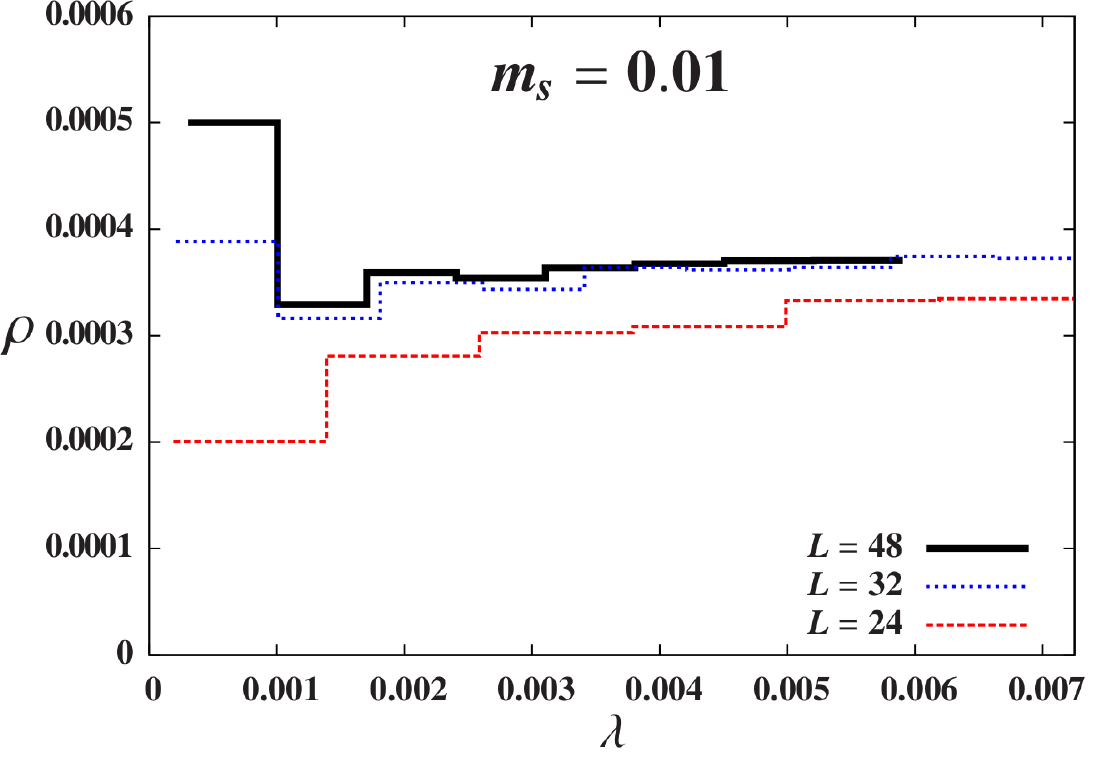}\hfill
  \includegraphics[width=0.45\linewidth]{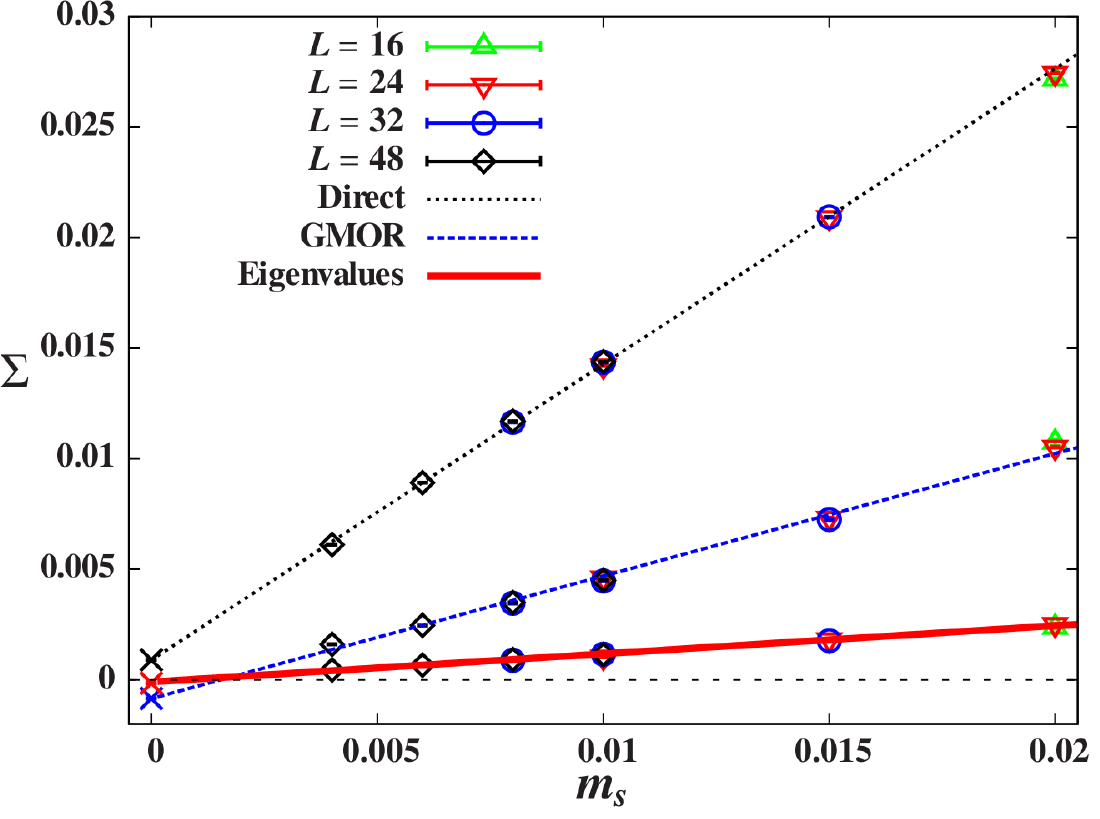}
  \caption{\label{fig:pbp} Left: Dirac eigenvalue density $\rho(\la)$ for ensembles with $m_s = 0.01$.  Right: The chiral condensate (normalized per continuum flavor) from direct measurements, the Gell-Mann--Oakes--Renner relation, and the eigenmode number, with chiral extrapolations using the six points listed in \protect\eq{eq:fits}.}
\end{figure}

In the chiral limit, the chiral condensate $\Si = \lim_{m \to 0}\pbp$ is the order parameter of S$\chi$SB.
We can probe \Si in three ways, all of which are shown in the right panel of \fig{fig:pbp}, normalized per continuum flavor.
Direct measurements of \pbp are dominated by a term $\propto m / a^2$, requiring a long extrapolation to the chiral limit.
Using the spectrum results discussed above, we apply the Gell-Mann--Oakes--Renner (GMOR) relation $\pbp = M_P^2 F_P^2 / (2m)$ to obtain another estimate that turns out to be less sensitive to $m$.

Finally, the Banks--Casher relation $\Si = \lim_{\la \to 0}\ \lim_{m\to 0}\ \lim_{V\to \infty} \pi\rho(\la)$ allows us to extract the chiral condensate from the eigenvalue density of the massless Dirac operator.
The left panel of \fig{fig:pbp} shows representative data for $\rho(\la)$, on different volumes with fixed dynamical fermion mass $m_s = 0.01$.
There are clear finite-volume effects in the $24^3\X48$ results (which we already saw in \fig{fig:Edin}), but $\rho(\la)$ for the larger volumes is roughly constant over a range of $\la$.
We estimate this constant by calculating the slope of its integral, the eigenmode number $\nu(\la) \propto \int_0^{\la} \rho(\om) d\om$. 
We have not attempted to remove any near-zero modes associated with the fluctuating topological charge shown in \fig{fig:therm}; for now we simply omit the first few bins of $\rho(\la)$ from our analysis.

The chiral extrapolations in the right panel of \fig{fig:pbp} all produce very small $\Si$, the largest being $\Si_{dir} = 9(2)\times10^{-4}$ from direct measurements.
In fact, when extrapolated linearly in $m_s$, the GMOR and eigenvalue results both lead to negative $\Si$.
Although \fig{fig:pbp} shows $m_s \to 0$ extrapolations, linear fits in terms of $M_P^2$ are much more stable for the GMOR and eigenvalue results (but not for the direct measurements or $M_V$ and $F_P$ discussed above).
From such $M_P^2 \to 0$ extrapolations, we find $\Si_{eig} = 0.55(5)\times10^{-4}$, while $\Si_{GMOR} = -1.5(7)\times10^{-4}$ is still slightly negative.
As in the previous section, these uncertainties are dominated by dependence on the fit range.
The points listed in \eq{eq:fits} correspond to $0.016 < M_P^2 < 0.11$, while omitting the lightest and heaviest points gives $0.025 < M_P^2 < 0.073$.
In short, we cannot reliably resolve a non-zero chiral condensate, even using $48^3\X96$ ensembles with fermion masses as light as $m = 0.004$ ($M_P = 0.13$).

\mysection{\label{sec:conclusion}Additional analyses, prospects and next steps} 
The results discussed above only scratch the surface of the physics potential in the eight-flavor ensembles being generated by USBSM.
Already we see that this system has a very small or vanishing chiral condensate, and the light hadron spectrum correspondingly shows no clear signs of spontaneous chiral symmetry breaking.
We identify where finite-volume effects are under control, and observe well-behaved thermalization and auto-correlations in the configuration generation.

In addition, we are carrying out valence domain wall measurements on ensembles with $m \geq 0.008$, to study the electroweak $S$ parameter following the methods of \refcite{Appelquist:2010xv}.
We have also developed unitary staggered measurements of the vector and axial-vector correlation functions that determine $S$, to see if the computational cost of valence domain wall measurements is truly justified.
The scalar spectrum is another high priority that we are actively investigating, with the extended $24^3\X48$ and $32^3\X64$ ensembles our first targets for glueball and disconnected diagram calculations.

Intrepid will retire at the end of 2013, by which time we hope to complete all of our $48^3\X96$ ensembles, and obtain thermalized configurations for at least one $64^3\X128$ run.
Given our current results for the spectrum and $\Si$, especially the preliminary value $F = 0.021(2)$, such a $64^3\X128$ ensemble will be extremely valuable in our attempt to access the chiral regime.

\mutesection{Acknowledgments} 
I thank the participants in the USBSM community for collaboration and review of this contribution.
In particular, George Fleming, Anna Hasenfratz, Meifeng Lin, James Osborn and Ethan Neil made important contributions to the work summarized above, which also benefited from conversations with Kieran Holland and Julius Kuti.
Part of this work was performed at the Aspen Center for Physics (NSF Grant No.~1066293), which I thank for support and hospitality.
Partial support came from DOE Grant Nos.~DE-SC0010005, DE-FC02-12ER41877 and DE-FG02-85ER40231. 
In addition to gauge generation on Intrepid, numerical analyses were carried out on clusters at Argonne, Livermore and Fermilab, the last managed by USQCD with support from the DOE.

\bibliographystyle{utphys}
\bibliography{Lattice13}
\end{document}